\documentstyle[epsfig]{mn}

\def\i{\item}
\def\bi{\bibitem{}}

\def\ni{\noindent}
\def\beb{\begin{thebibliography}{999}}
\def\eeb{\end{thebibliography}}
\def\bei{\begin{itemize}}
\def\eei{\end{itemize}}
\def\bef{\begin{figure}}
\def\eef{\end{figure}}
\def\ben{\begin{enumerate}}
\def\een{\end{enumerate}}
\def\beq{\begin{equation}}
\def\eeq{\end{equation}}
\def\ber{\begin{eqnarray}}
\def\eer{\end{eqnarray}}
\def\edo{\end{document}}

\newcommand{\dmdt}{{\mbox{{\rm M}$_{\odot}$}} {\rm yr}$^{-1}$}

\newcommand{\gcc}{{\rm g} \, {\rm cm}^{-3}}

\newcommand{\msun}{\mbox{{\rm M}$_{\odot}$}}
\newcommand{\mdot}{\mbox{$\dot{M}$}}

\newcommand{\lsim}{\raisebox{-0.3ex}{\mbox{$\stackrel{<}{_\sim} \,$}}}

\begin{document}
\title[neutron star field evolution]
{Magnetic Field Evolution of Accreting Neutron Stars - II}
\author[Konar and Bhattacharya]
{Sushan Konar$^{1,2,{\ast},{\dag}}$ and Dipankar Bhattacharya$^{1,{\dag}}$ \\ 
$^1$Raman Research Institute, Bangalore 560080, India \\
$^2$Joint Astronomy Program, Indian Institute of Science, Bangalore 560012, India \\
$^{\ast}$Present Address : IUCAA, Pune 411007, India \\ 
$^{\dag}$ e-mail : sushan@iucaa.ernet.in, dipankar@rri.ernet.in \\ }
\date{21st April, 1998}
\maketitle

\begin{abstract}
We investigate the evolution of the magnetic field of isolated pulsars and of neutron stars in 
different kinds of binary systems, assuming the field to be originally confined to the crust. 
Our results for the field evolution in isolated neutron stars helps us to constrain the physical 
parameters of the crust. Modelling the full evolution of a neutron star in a binary system through 
several stages of interaction we compare the resulting final field strength with that observed in 
neutron stars in various types of binary systems. One of the interesting aspects of our result is a 
positive correlation between the rate of accretion and the final field strength, for which some 
observational indication already exists. Our results also match the overall picture of the field 
evolution in neutron stars derived from observations. 
\end{abstract}


\section{introduction}
\label{sintro}

\ni The astronomical objects that have so far been unambiguously identified with neutron stars can be divided 
into two distinct classes - a) the radio pulsars and b) the X-ray binaries containing a neutron star. Observations 
suggest that a subset of radio pulsars also descend from X-ray binaries. And neutron stars in X-ray binaries 
themselves represent an evolutionary state beyond that of isolated radio pulsars (see Bhattacharya 1995b, 1996a, 
van den Heuvel 1995, Verbunt \& van den Heuvel 1995 and references therein). The present belief is that this 
evolutionary link can be understood within an unified picture of evolution of their spin as well as the magnetic 
field. The spin-evolution of the neutron stars in binary systems has been investigated following the detailed binary 
evolution assuming some simple model for the field evolution (Verbunt \& van den Heuvel 1995 and references therein). 
But an integrated picture of the spin and microscopic field evolution models has not been investigated in detail. 
Nevertheless, from observational facts and from recent statistical analyses made on the pulsar population, the 
following conclusions can be drawn regarding the field evolution.
\begin{itemize}
\i Isolated pulsars with high magnetic fields ($ \sim 10^{11}$ -- $10^{13}$ G) do not undergo any 
significant field decay during their lifetimes (Bhattacharya et al.\/ 1992; Wakatsuki et al.\/ 1992, 
Lorimer 1994, Hartman et al. 1997).
\i The fact that binary pulsars, as well as millisecond and globular cluster pulsars which almost always 
have a binary history, possess much lower field strengths suggests that significant field decay occurs 
only as a result of the interaction of a neutron star with its binary companion (Bailes 1989).
\i The age determination of pulsars with white dwarf companions show that most of the millisecond pulsars 
are extremely long-lived. The old age ($\sim 10^9$ years) of the low-field pulsars implies that their field 
is stable over long time-scales, i.e., after the recycling process in binaries is over, the field does 
not undergo further decay (Bhattacharya \& Srinivasan 1986; Kulkarni 1986; van den Heuvel, van Paradijs 
\& Taam 1986; Verbunt et al.\/ 1990). 
\i The binary pulsars with massive companions, like the Hulse-Taylor pulsar, have field strengths in excess 
of $10^{10}$~Gauss, whereas the low mass binary pulsars include both high-field pulsars and very low-field 
objects like the millisecond pulsars. The present belief regarding the evolutionary history of the binary 
pulsars is that the high mass binary pulsars come from systems similar to the high mass X-ray binaries (HMXBs)
whereas the low mass binaries are the progenies of the low mass X-ray binaries (LMXBs). This indicates that 
the nature of field evolution is affected by the kind of companion the neutron star has.
\i The evolutionary link between millisecond pulsars and low-mass X-ray binaries seem to be borne out both 
from binary evolution models and from the comparative study of the kinematics of these two populations. To 
spin a neutron star up to millisecond periods at least an amount $\sim 0.1 \msun$ needs to be accreted. Such 
huge amount of mass transfer is possible only in LMXBs, where the duration of mass transfer could be as long 
as $10^9$ years. Even though the present estimates for the birthrate of LMXBs fall short of that of the 
millisecond pulsars (Kulkarni \& Narayan 1988; Lorimer 1995), kinematic studies (Cordes et al. 1990; Wolszcan 
1994; Nice \& Taylor 1995; Nicastro \& Johnston 1995; Bhattacharya 1996a, Ramachandran \& Bhattacharya 1997; 
Cordes \& Chernoff 1997) indicate that the two populations are most likely to be related as both have very 
similar kinematic properties. Moreover, the recent observation of the 2.49-millisecond X-ray pulsar SAX 
J1808.4-3658 (Wijnands \& van der Klis 1998) with an estimated dipole field strength of 
$\sim 2 - 6 \times 10^8 ~ G$ is a direct pointer to the connection between LMXBs and millisecond pulsars. 
It is understood that this object would emerge as a typical millisecond radio pulsar when the mass accretion 
in the system stops, vindicating present theoretical expectations. Besides the birthrate problem there is no 
satisfactory explanation for the origin of {\em isolated} millisecond pulsars either. Any model for field 
evolution has to be consistent with the nature of binary evolution that produced these objects. The field 
evolution scheme must also provide for a limiting minimum field strength---the so-called `flooring' seen at 
$\sim 10^8$~G should arise out of the evolution itself. Except for a few of the models (Romani 1995, Jahan Miri 
\& Bhattacharya 1994, Konar \& Bhattacharya 1997) an explanation for this has not been attempted so far.  
\end{itemize}

\ni There are two classes of models which have been explored in this context: one that relates the magnetic 
field evolution to the spin evolution of the star and the other attributing the field evolution to direct 
effects of mass accretion (see Ruderman (1995) and Bhattacharya (1995a) for detailed reviews). Almost all 
the models are built upon two themes, namely, a large scale macroscopic restructuring of the fields in the 
interior of the star (for example, the spin-down and expulsion of flux from the superfluid core of the star) 
and a microscopic mechanism (like ohmic dissipation of the currents in the crust) being effective to actually 
dissipate the underlying currents supporting the observable field. The different class of models usually assume 
different kind of initial field configurations. Models depending on spin-down assume a core-flux supported by 
proton superconductor flux tubes. Whereas, models invoking ohmic dissipation usually assume an initial crustal 
configuration. The mechanism of ohmic decay, being unique to the crustal currents, is also used in models where 
spin-down is invoked for flux expulsion, for a subsequent dissipation of such flux in the crust (Jahan Miri \& 
Bhattacharya 1994, Bhattacharya \& Datta 1996). \\

\ni In an accretion-heated crust the decay takes place principally as a result of rapid dissipation of currents due 
to the decrease in the electrical conductivity and hence a reduction in the ohmic dissipation time-scale (Geppert \& 
Urpin (1994), Urpin \& Geppert (1995), Urpin \& Geppert (1996), Konar \& Bhattacharya (1997) - paper-I from now on). 
The crustal field undergoes ohmic diffusion due to the finite electrical conductivity of the crustal lattice, but 
the time-scale of such decay is very long under ordinary conditions (Sang \& Chanmugam 1987; Urpin \& Muslimov 1992).
The situation changes significantly when accretion is turned on. The heating of the crust reduces the electrical 
conductivity by several orders of magnitude, thereby reducing the ohmic decay time-scale. There is also an additional 
effect that acts towards stabilizing the field. As the mass increases, a neutron star becomes more and more compact 
and the mass of the crust actually decreases by a small amount. So the newly accreted material forms the crust and 
the original crustal material gets continually assimilated into the superconducting core below. The original current 
carrying layers are thus pushed into deeper and more dense regions as accretion proceeds. The higher conductivity of 
the denser regions would progressively slow down the decay, till the current loops are completely inside the 
superconducting region where any further decay is prevented (paper I). \\

\ni In paper I we have discussed a model for the evolution of the magnetic field in accreting neutron stars, 
assuming an initial crustal flux. It has been borne out by our calculations that the model is capable of 
explaining most of the characteristics needed to explain the observed low-field pulsars. In this work, we 
confront the field evolution model of paper-I, namely that of assuming an initial crustal flux, with 
observations of both isolated neutron stars and neutron stars in binary systems. In particular, we address 
the question of millisecond pulsar generation purely from the point of view of the field evolution. Our results 
are consistent with the general view that millisecond pulsars come from low mass x-ray binaries. We also find 
that neutron stars processed in HMXBs would retain fairly high field strengths in conformity with what is 
observed in high mass radio pulsars. A recent work by Urpin, Konenkov \& Geppert (1998) has undertaken similar 
investigations for the case of HMXBs. Even though their calculations do not extend up to the late stages of 
binary evolution, our results match with theirs within the valid regime of comparison.

\section{physical processes in neutron star binaries}
\label{sphysics}

\ni From the point of view mass transfer, neutron stars in binaries, in general, go through three distinct 
phases of evolution: 
\ben
\i The Isolated Phase - Though the stars are gravitationally bound, there is no mass transfer. Therefore both 
the spin and the magnetic field evolve as they would in an isolated pulsar. The spin undergoes a pure dipole 
slow-down during this phase.
\i The Wind Phase - The interaction is through the stellar wind of the companion which is likely to be in its 
main-sequence. The wind phase may have two distinct sub-phases, namely - the propeller phase, in which the star 
spins down, and the phase of wind accretion (Verbunt 1993, Bhattacharya \& van den Heuvel 1991). The star may be 
spun up in the accretion phase, the maximum spin-up for a given rate of accretion and a given strength of the 
surface field, being given by the following relation (Alpar, Cheng, Ruderman \& Shaham 1982, Chen \& Ruderman 1993):
\beq
P_{\rm eq} = 1.9 \, {\rm ms} \; B_9^{6/7} \; (\frac{M}{1.4 \msun})^{-5/7} \; 
(\frac{\mdot}{\mdot_{\rm Edd}})^{-3/7} \; R_6^{18/7} \label{ePeq}
\eeq
where $B_9$ is the surface field in $10^9$ Gauss, $M$ is the mass of the star, $\mdot$ is the rate of 
accretion, $\mdot_{\rm Edd}$ is the Eddington rate of accretion and $R_6$ is the radius of the star in units 
of 10~km. The duration of these two sub-phases vary widely from system to system and the phase of actual 
wind accretion may not at all be realized in some cases. For models based on an initial crustal field 
configuration, though, the phase of wind accretion and the subsequent phase of Roche contact play the 
all-important role.
\i The Roche-contact Phase - When the companion of the neutron star fills its Roche-lobe a phase of heavy 
mass transfer ($\mdot \sim \mdot_{\rm Edd}$) ensues. Though short-lived in case of HMXBs ($\sim 10^4$ years), 
this phase can last as long as $10^9$ years for LMXBs. Consequently, this phase is very important for field 
evolution in LMXBs. 
\een

\ni The nature of the binary evolution is well studied for the LMXBs and the wind phase of the HMXBs where the 
mass transfer proceeds in a controlled manner. However, in HMXBs the Roche-lobe overflow at the end of the wind
phase may completely engulf the neutron star giving rise to a {\em common-envelope} phase. The exact nature of 
mass transfer in this phase is not known. But being short-lived this phase does not affect the field evolution 
significantly. Moreover, not much attention has been paid to the evolution of the intermediate systems with 
companion masses in the range $\sim$ 2 - 5 \msun. They are most likely to have an intermediate nature in that 
the wind phase is prolonged and the accretion rates are similar to those in the wind phase of low-mass systems, 
whereas, the Roche-contact phase is perhaps similar to that in HMXBs. In either of the phases it would be 
difficult to observe such systems. So far no intermediate-mass system has been observed in the X-ray phase. As 
for the pulsars processed in them, there are perhaps three examples -- PSR B0655+64, PSR J2145-0750 and 
PSR J1022+1001 (Camilo et al. 1996). But because of the uncertainties surrounding their mass transfer history 
we exclude this kind of binaries from the present discussion. 

\section{computations}
\label{scomp}

\ni The evolution of the magnetic field is governed by the equation 
\beq
\frac{\partial \vec B}{\partial t} = \vec \nabla \times (\vec V \times \vec B) - \frac{c^2}{4 \pi} \vec \nabla \times
(\frac{1}{\sigma} \times \vec \nabla \times \vec B),
\eeq
where $\vec V$ is the velocity of material movement and $\sigma$ is the electrical conductivity of the medium. We follow the mass-transfer 
history of neutron stars in high-mass and low-mass binary systems using the methodology adopted in paper-I to solve the above equation. 
The radius and the crustal mass of a neutron star remain effectively constant for the maximum amount of accreted masses considered, and 
the corresponding change in the crustal density profile is negligible. We therefore take the mass flux to be the same throughout the crust, 
equal to its value at the surface. Assuming the mass flow to be spherically symmetric in the crustal layers of interest, one obtains the 
velocity of material movement to be
\beq
\vec V = - \frac {\mdot}{4\pi r^2 \rho (r)} \hat{r},
\eeq
where \mdot\ is the rate of mass accretion and $\rho (r)$ is the density as a function of radius $r$. For the case of 
isolated pulsars we set $\vec V = 0$. \\

\ni It is evident from the above equations that for an assumed crustal flux in an isolated neutron star the field 
decreases due to pure ohmic dissipation of the current loops in the crust. The diffusive time-scale, through the 
electrical conductivity, is dependent on three factors, namely
\bei
\i the density at which the initial current loops are located,
\i the impurity content of the crust, and 
\i the temperature of the crust.
\eei
In an accreting neutron star the rate of mass accretion has a major impact on the nature of field evolution in 
addition to the factors mentioned above. 

\subsection{Crustal Physics}
\ni We have seen in paper-I that the field evolution stops as the field {\em freezes in} when about 10\% 
of the original crustal mass is accreted. This happens due to the fact that by then the current loops reach 
regions of extremely high electrical conductivity. The mass of the crust of a 1.4 $\msun$ neutron star, 
with our adopted equation of state (Wiringa, Fiks \& Fabrocini (1988) matched to Negele \& Vautherin (1973) 
and Baym, Pethick \& Sutherland (1971) for an assumed mass of 1.4~\msun), is $\sim 0.044 \msun$. Therefore,
in the present investigation we stop our evolutionary code when $\sim 0.01 - 0.04~\msun$ is accreted. From our 
results (presented in section [\ref{sresults}]) it is evident that the field evidently shows signs of 
{\em freezing in} when we stop our calculation. Of course, as we have mentioned earlier, to achieve spin-up to
millisecond periods about 0.1 solar mass requires to be accreted. But once the field attains its `frozen-in' 
value subsequent accretion does not affect it. Therefore, the final spin-period is determined by this 
value of the surface field in accordance with equation (\ref{ePeq}). 

\subsection{Thermal Behaviour}
\ni An isolated neutron star cools down after birth chiefly by copious neutrino emission. It has recently been 
shown (see Page, 1998 and references therein for details) that there are many uncertainties in this regard. 
Existing observational data could be made to fit both the `standard' and the `accelerated' cooling with 
appropriate assumptions regarding the state of the stellar interior. Therefore, for the sake of completeness, 
we have looked at both the scenarios. The actual behaviour of the system is most likely to be something in 
between. We have used data from V.~A. Urpin \& K.~A. van Riper (private communications) for both standard and 
accelerated cooling. The thermal behaviour of isolated pulsars and of neutron stars accreting material from their 
binary companions differ from each other significantly (Page, 1997). For a neutron star that is a member of a 
binary, the thermal behaviour will be similar to that of an isolated neutron star until mass accretion on its 
surface begins. Therefore through the isolated and the propeller phase the neutron star cools like an isolated 
one. In particular, in the low-mass systems the duration of the isolated phase and the propeller phase could 
be quite long and therefore is rather important in affecting the subsequent evolution of the surface field. We 
have shown in paper-I that a phase of field evolution in the isolated phase modifies the subsequent evolution 
considerably. Therefore, it is necessary to take into account the proper cooling history of the neutron star 
prior to the establishment of contact with its binary companion. When, in the course of binary evolution, the 
neutron star actually starts accreting mass - the thermal behaviour changes from that characteristic of an 
isolated phase. The crustal temperature then settles down to a steady value determined by the accretion rate 
as given by the following fitting formula (see paper-I):
\beq
\log T = 0.397 \log \mdot + 12.35.           \label{etdmdt}
\eeq

\subsection{Binary Parameters}
\ni The binary evolution parameters for the LMXBs and the HMXBs used by us are as follows (Verbunt 1993; 
Bhattacharya \& van den Heuvel 1991; van den Heuvel 1992; van den Heuvel \& Bitzaraki 1995; King et al. 1995) -
\ben
\i Low Mass X-ray Binaries - 
\bei
\i Isolated phase - Though binaries with narrow orbits may not have a long-lived phase of completely detached 
evolution, long period binaries (like the progenitor system of PSR 0820+02 with $\sim$ 250 day orbital period 
(Verbunt \& van den Heuvel 1995)) may spend longer than $10^9$ years before contact is established. In general, 
the isolated phase lasts between $10^8 - 10^9$ years.  
\i Wind phase - This phase again lasts for about $10^8 - 10^9$ years with attendant rates of accretion 
ranging from about $10^{-15}$~\dmdt to $10^{-12}$~\dmdt (paper-I).
\i Roche-contact phase - In this phase, the mass transfer rate could be as high as the Eddington rate 
($10^{-8}$~\dmdt for a 1.4 $\msun$ neutron star), lasting for $\lsim 10^8$ years. But there has been recent 
indications that the low-mass binaries may even spend $\sim 10^{10}$ years in the Roche-contact phase with a 
sub-Eddington accretion rate (Hansen \& Phinney 1997). For wide binaries, however, the contact phase may last 
as little as $10^7$ years. We have investigated the cases with accretion rates of $10^{-10}$~\dmdt and 
$10^{-9}$~\dmdt. With a higher accretion rate the material movement is faster and therefore the `freezing-in' 
takes place earlier. Moreover, the equation (\ref{etdmdt}), used by us to find the crustal temperature for a 
given rate of accretion, gives temperatures that are too high for accretion rates above $10^{-10}$~\dmdt. The 
neutrino cooling is likely to prevent the temperature from reaching such large values and therefore for high 
rates of accretion the crustal temperature would probably reach a saturation value. In our calculations 
we assume that the temperature that could be attained by accretion-induced heating is $\sim 10^{8.5}$~K for an 
accretion rate of $10^{-9}$~\dmdt. Even this is probably an overestimate of the crustal temperature in an
accreting neutron star. In view of this we have restricted our calculation to the above-mentioned rates of 
accretion and have not investigated the cases for near-Eddington to super-Eddington rates.
\eei
\i High Mass X-ray Binaries - 
\bei
\i Isolated phase - This phase is short in binaries with a massive companion and may last as little as ten 
thousand years.
\i Wind phase - This phase is also relatively short (compared to the low-mass systems), lasting not more 
than $10^7$ years (equivalent to the main-sequence life time of the massive star), with accretion rates ranging 
from $10^{-14}$~\dmdt to $10^{-10}$~\dmdt. 
\i Roche-contact phase - A rapid phase of Roche-lobe overflow follows the wind phase. The rate of mass 
shedding by the companion could be as high as one tenth of a solar mass per year, of which a tiny fraction 
is actually accreted by the neutron star (the maximum rate of acceptance being presumably equal to the Eddington 
rate). The duration of this phase is $\lsim 10^4$ years.
\eei
\een
We must mention here about the Globular Cluster binaries though we have not performed any explicit calculations 
for such systems. The duration and the rates of mass accretion vary widely depending on how the mass transfer 
phase ends in such binaries. The mass transfer may end through binary disruption, by the neutron star spiraling 
in or through a slow turn off. 

\section{results and discussions}
\label{sresults}

\subsection{Solitary Neutron Stars}

\bef
\begin{center}{\mbox{\epsfig{file=plot_diffusion_q1.ps,width=150pt,angle=-90}}}\end{center}
\caption[]{The evolution of the surface magnetic field due to pure diffusion. Curves 1 to 5 correspond to densities 
of $10^{13}, 10^{12.5}, 10^{12}, 10^{11.5}, 10^{11}~\gcc$ respectively, at which the initial current configurations 
are centred. Standard cooling has been assumed for the solid curves and an accelerated cooling for the dotted ones. 
Notice that for accelerated cooling, curves 1 through 4 are almost indistinguishable. All curves correspond to $Q$ = 0.}
\label{fdiffusion_q1}
\eef

\bef
\begin{center}{\mbox{\epsfig{file=plot_diffusion_q2.ps,width=150pt,angle=-90}}}\end{center}
\caption[]{Same as figure [\ref{fdiffusion_q1}] but with $Q$ = 0.05. Only the case with standard cooling
has been presented here.}
\label{fdiffusion_q2}
\eef

\ni Figure [\ref{fdiffusion_q1}] shows the evolution of the surface field in an isolated neutron star due to 
pure ohmic decay, for different densities at which the initial current distribution is concentrated and for different 
assumed cooling behaviour. The lower the density of current concentration the faster the field decays. This figure 
refers to a situation where the impurity strength, $Q$ (see paper I), has been assumed to be zero. In figure 
[\ref{fdiffusion_q2}] similar curves for the evolution of the surface field with an assumed $Q$ = 0.05 are plotted.  
In this case the surface field shows a trend towards sharp decline even in the later phases of evolution, unlike
in figure [\ref{fdiffusion_q1}]. Statistical analyses indicate that the surface fields of solitary pulsars undergo 
insignificant decay between the ages of $10^5$ - $10^8$ years (Bhattacharya et al.\/ 1992; Wakatsuki et al.\/ 1992, 
Lorimer 1994, Hartman et al. 1997).  
Therefore such large impurities in the crust are ruled out. Our calculations show that the maximum permissible value
of $Q$, consistent with the indications from the statistical analyses, is $\sim 0.01$. From the curves corresponding to 
a faster cooling mechanism in figure [\ref{fdiffusion_q1}] it can be seen that the actual reduction in the field strength
is much less. However, the nature of the decay is quite similar to the case with standard cooling. Consequently, the 
maximum possible value of $Q$ remains the same irrespective of the nature of cooling assumed and provides for an
absolute limit.

\subsection{High Mass Binaries}

\ni Figure [\ref{fhmxb}] shows the evolution of the surface field in the HMXBs; for two different values of the 
density at which the initial current profile is centred. A standard cooling has been assumed for the initial
isolated phase of the neutron star. Calculations show that an assumption of an accelerated cooling in the isolated
phase does not change the final results significantly. We assume a Roche-contact phase lasting for $10^4$ years 
with an uniform rate of accretion of $10^{-8}$~\dmdt. Since this phase is extremely short-lived, the actual field 
decay takes place during wind accretion. In fact, the decay attained in the Roche-contact phase is insignificant as 
can be seen from the last portion of the figure [\ref{fhmxb}] where this phase is plotted in an expanded scale. \\

\bef
\hspace{-0.4cm}
\mbox{\epsfig{file=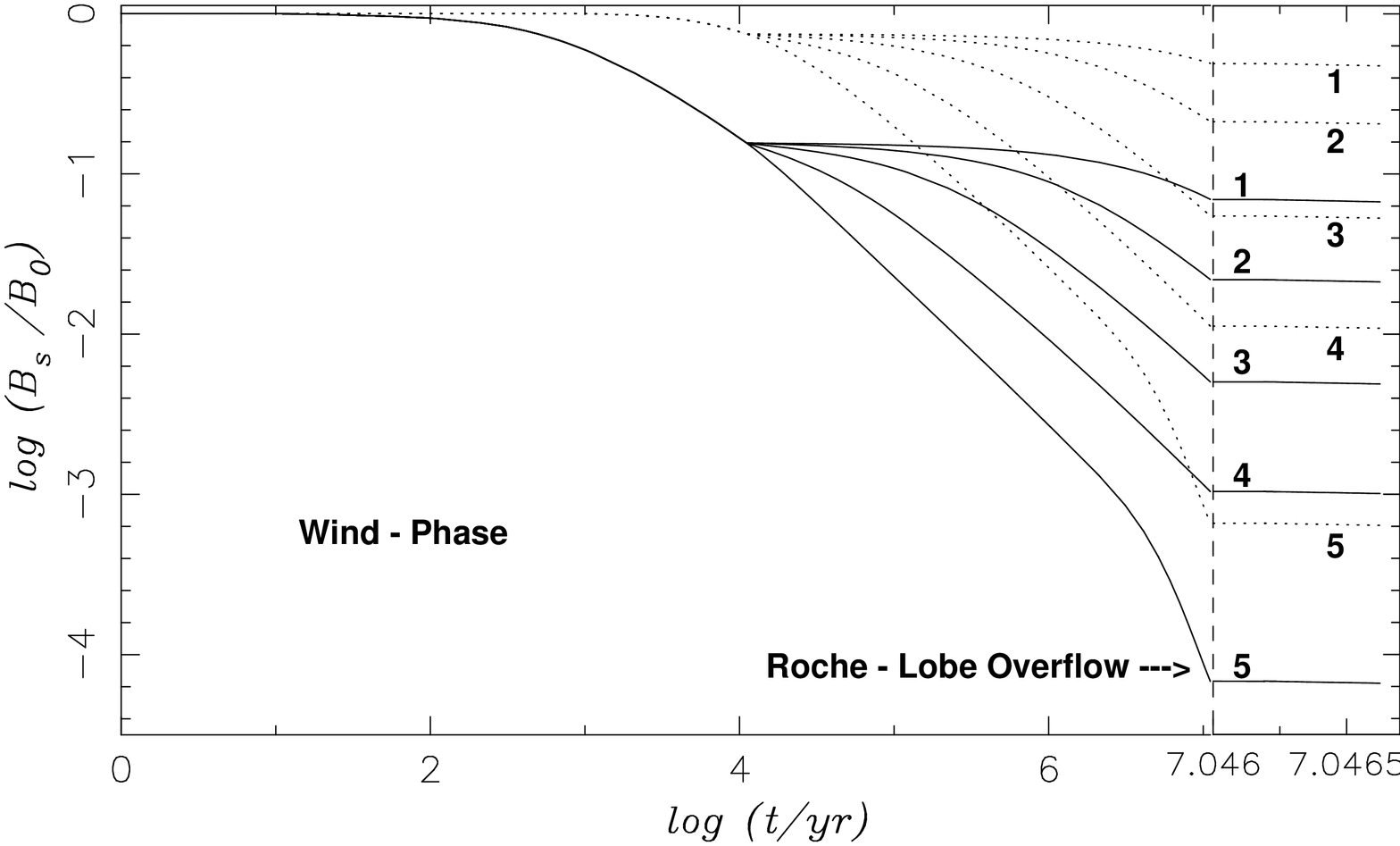,width=170pt,angle=-90}}
\caption[]{Evolution of the surface magnetic field in HMXBs for four values of wind accretion rate. The solid and the dotted 
curves correspond to initial current configurations centred at $\rho = 10^{11}, 10^{13}~\gcc$ respectively. Curves 1 to 5 
correspond to $\mdot = 10^{-14}, 10^{-13}, 10^{-12}, 10^{-11}, 10^{-10}$~\dmdt. For all curves an accretion rate of 
$\mdot = 10^{-8}$~\dmdt in the Roche-contact phase and $Q$ = 0.0 have been assumed. The Roche-contact phase has been plotted in 
an expanded scale.}
\label{fhmxb}
\eef

\ni It is evident that if the initial current distribution is located at high densities
the objects from high-mass systems will retain fairly large final fields. Even for current distributions 
located at lower densities if the duration of the wind phase is not too long - high-field objects are 
produced.  We expect these objects to show up as recycled pulsars with relatively high fields and long 
periods like PSR B1913+16 or PSR B1534+12. On the other hand, if the wind phase lasts for about $10^7$ years, 
it is possible to obtain a significant field decay for higher rates of accretion in that phase. But as the total 
mass and hence the total angular momentum accreted is not sufficient to spin the star up to very short periods, 
these systems probably would not produce millisecond pulsars. In other words these so-called `recycled' pulsars 
would have small magnetic fields with relatively long spin-periods and therefore may not at all be active as 
pulsars. We make an estimate of the actual spin-up for objects processed in high-mass systems to check this fact.\\

\ni The spin-up of a neutron star, in a binary system, is caused by the angular momentum brought in by the 
accreted matter. In magnetospheric accretion matter accretes with angular momentum specific to the Alfv\'{e}n 
radius. Therefore, the total angular momentum brought in by accretion is
\beq
J_{\rm accreted} = \delta M R_A V_A
\eeq
where $\delta M$ is the total mass accreted. $R_A$ and $V_A$ are the Alfv\'{e}n radius and Keplerian velocity at 
that radius. The final period of the neutron star then is 
\beq
P_{\rm final} = 2 \pi \frac{I_{\rm ns}}{J_{\rm accreted}},
\eeq
where $I_{\rm ns}$ is the moment of inertia of the neutron star. In figure [\ref{fhmxb_spinup}] we have indicated 
the possible location of the `recycled' pulsars originating in HMXBs. We find that for sufficiently low field 
strengths ($B \lsim 10^9$~Gauss) the spun-up neutron stars indeed will be beyond the death line. And active pulsars 
with slightly higher fields will lie very close to the death line. Therefore, even though spun-up neutron stars will 
be produced in the entire region right of the dashed line (corresponding to a total amount of mass accreted) the
only ones active as pulsars would be the ones with relatively high magnetic field. As can be seen from figure
[\ref{fhmxb_spinup}] the minimum field for a recycled pulsar from an HMXB is $\sim 10^{9.4}$~G for a total accreted mass
of $10^{-3}$~\msun and $\sim 10^{10.6}$~G for a total accreted mass of $10^{-4}$~\msun. The exact estimate of low-field
recycled pulsars from HMXBs could be obtained from detailed population synthesis investigation of these systems.\\

\ni It should be noted here that even though the nature of field evolution is very different in the isolated phase 
for standard and accelerated cooling, the final surface field values at the end of the wind phase are not very 
different. This is due to the fact that the nature of the field evolution is significantly influenced by previous 
history. We have seen in paper-I that subsequent decay is slowed down in a system with a history of prior field decay 
than in systems without any. The decay in the isolated phase is less for accelerated cooling, but in the subsequent 
wind accretion phase the field decays more rapidly in such systems than in those starting with standard cooling.

\bef
\begin{center}{\mbox{\epsfig{file=plot_hmxb_spinup.ps,width=150pt,angle=-90}}}\end{center}
\caption[]{The probable location of HMXB progenies in the field-period diagram. The dashed lines correspond to the 
maximum spin-up achievable, the upper and the lower lines being for assumed accreted masses of $10^{-4}$ and 
$10^{-3}$~\msun. The recycled pulsars from HMXBs are expected to lie within the hatched region.}
\label{fhmxb_spinup}
\eef
     
\subsection{Low Mass Binaries}

\ni Figure [\ref{flmxb}] shows the evolution of the surface field in LMXBs. In this figure we plot the complete evolution 
for two values of the accretion rate in the Roche-contact phases and five values of the density at which currents are 
initially concentrated, assuming standard cooling for the isolated phase. Both the wind and the Roche-contact phase are 
plotted in expanded scales to highlight the nature of the field evolution in these phases. As in the case of the HMXBs, 
the assumption of a particular cooling scenario (standard or accelerated) does not affect the final values of the surface 
field. \\

\bef
\begin{center}{\mbox{\epsfig{file=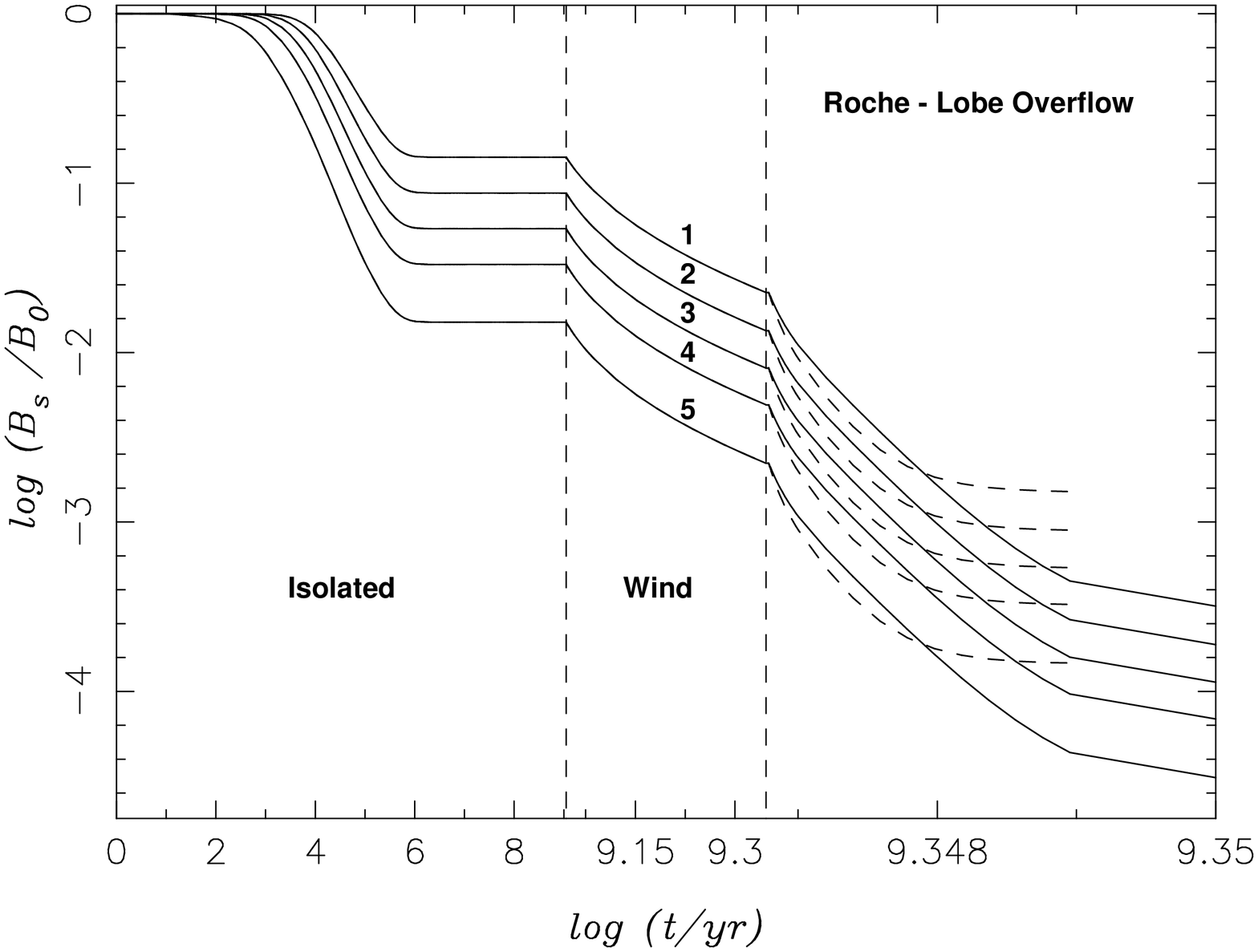,width=200pt,angle=-90}}}\end{center}
\caption[]{Evolution of the surface magnetic field in LMXBs with an wind accretion rate of $\mdot = 10^{-14}$~\dmdt. Curves 1 to 5 
correspond to initial current configurations centred at $\rho = 10^{13}, 10^{12.5}, 10^{12}, 10^{11.5}, 10^{11} \gcc$. All curves 
correspond to $Q$ = 0.0. A standard cooling has been assumed for the isolated phase. The wind and Roche-contact phases are plotted 
in expanded scales. The dashed and the solid curves correspond to accretion rates of $\mdot = 10^{-9}, 10^{-10}$~\dmdt in the Roche-contact phase.}
\label{flmxb}
\eef

\ni The surface field drops by half to one order of magnitude (depending on the rate of accretion) in the wind phase of 
the binary evolution. When the system is in contact through Roche-lobe overflow the field decay depends very much on 
the rate of accretion. A difference in the accretion rate in this phase shows up as a difference in the final value of 
the surface field, which freezes at a higher value for higher rates of accretion. The total decay in the Roche-contact 
phase may be as large as two orders of magnitude with respect to the magnitude of the field at the end of the wind phase. \\

\ni We have mentioned before that the phase of wind accretion may not be realized in some of the systems at all. In figure 
[\ref{flmxb_nw}] we have plotted the evolution of the surface field for such cases. We find that the final field strengths 
achieved without a phase of wind accretion is not very different from the cases where such a phase does exist. This is 
again indicative of the fact that a prior phase of ohmic diffusion slows down the decay in the subsequent phase and 
therefore the final result from both the cases tend to be similar. \\

\bef
\begin{center}{\mbox{\epsfig{file=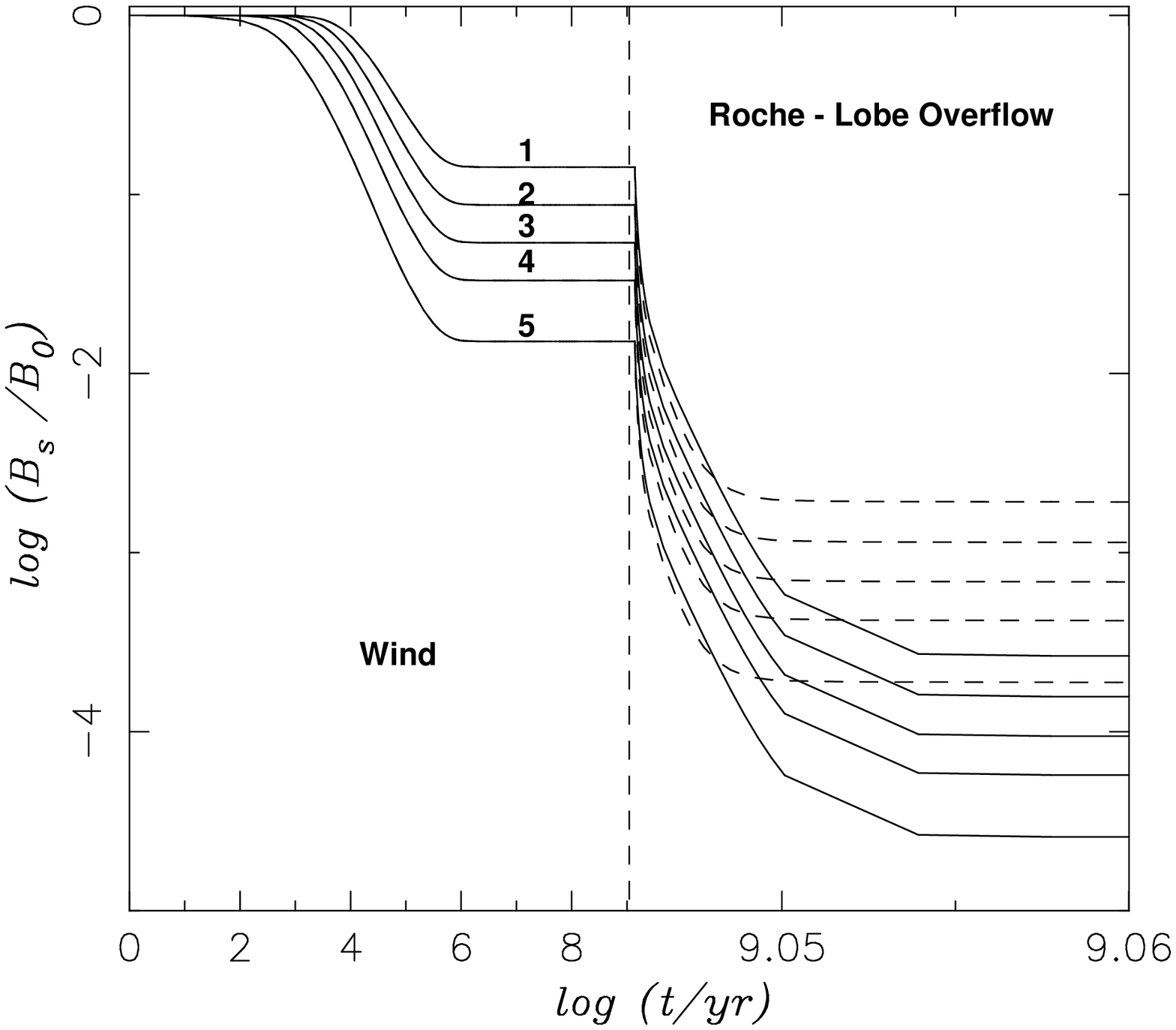,width=225pt,angle=-90}}}\end{center}
\caption[]{Evolution of the surface magnetic field in LMXBs without a phase of wind accretion and accretion rates of 
$\mdot = 10^{-10}, 10^{-9}$~\dmdt (the solid and dashed curves respectively) in the Roche-contact phase. Curves 1 to 
5 correspond to initial current configurations centred at $\rho = 10^{11}, 10^{11.5}, 10^{12}, 10^{12.5}, 10^{13} \gcc$. 
All curves correspond to $Q$ = 0.0. A standard cooling has been assumed for the isolated phase. The Roche-contact 
phase has been shown in expanded scale.}
\label{flmxb_nw}
\eef

\ni There are several interesting points to note here. Figures [\ref{flmxb}] and [\ref{flmxb_nw}] show that for higher values of 
accretion rate in the Roche-contact phase the final field values are higher. We have not explored the case of accretion with an 
Eddington rate in this phase for reasons mentioned before. From the trends observed in our calculation it is evident that with such 
a high rate of accretion the final field value may remain fairly large (see paper-I). Under such circumstances it will be possible 
to have `recycled' pulsars of high surface magnetic field (and therefore long spin-period) from low mass binaries and pulsars like 
PSR 0820+02 will fit the general scenario quite well. Then, of course, we do find significant amount of field decay with lower rates 
of accretion in the Roche-contact phase. Such low surface fields combined with the provision of maximal spin-up would produce 
millisecond pulsars. Therefore it is evident that the model of field evolution assuming an initial crustal field configuration is 
quite consistent with the present scenario of field evolution. The LMXBs will produce high-field, long-period pulsars in addition to 
the expected crop of millisecond pulsars. Whereas the only kind of recycled pulsars that are expected from the HMXBs would be of the 
relatively high-field, long-period variety. \\

\bef
\begin{center}{\mbox{\epsfig{file=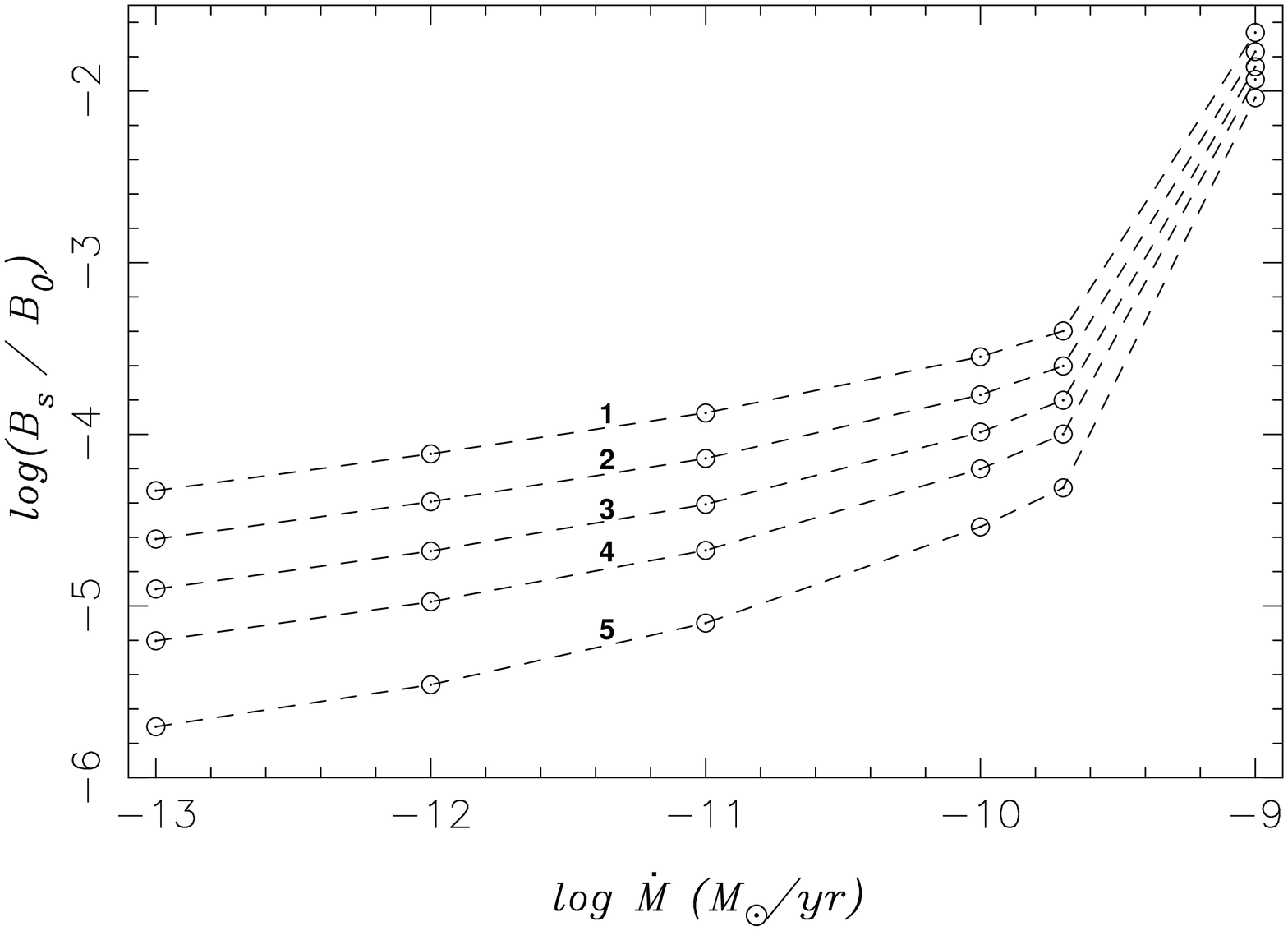,width=175pt,angle=-90}}}\end{center}
\caption[]{Final residual surface field vs. rate of accretion. Curves 1 to 5 correspond to initial current configurations centred at 
$\rho = 10^{13}, 10^{12.5}, 10^{12}, 10^{11.5}, 10^{11} \gcc$. }
\label{fresidual_field}
\eef

\ni It is clearly seen from the above results that there is a positive correlation between the rate of accretion with that of the final 
field strength, namely, the higher the rate of accretion the higher is the final field. This correlation, of course, refers to an average 
rate of accretion over the initial accreted mass of $\sim 0.01$~\msun. Accretion taking place later, whatever rate it might have, does not 
change the residual field. Figure [\ref{fresidual_field}] clearly shows the above-mentioned correlation. \\

\ni There has already been mention of such a correlation in connection with the Z and Atoll sources (Hasinger \& van 
der Klis 1989). It was suggested that the difference between these two classes of sources in regard to their fluctuation 
spectra is not only due to a difference in the accretion rate but also due to a difference in the magnetic field. And 
there has been indication from the study of the radiation spectra of these sources that the accretion rate and the 
magnetic field strength are positively correlated. Recently, Psaltis \& Lamb (1997) on the basis of LMXB spectra and 
White \& Zhang (1997) on the basis of the properties of the kilohertz QPOs have indicated the existence of such a 
correlation. \\

\ni There is another indirect evidence of the correlation between the rate of accretion and the strength of the 
surface magnetic field. From the data published by van den Heuvel \& Bitzaraki (1995) for binary pulsars, with low 
mass companions, it is clearly seen that the orbital period and the magnetic field are positively correlated. It is 
expected from evolutionary models (eg. Webbink, Rappaport \& Savonije 1983) that LMXB systems with longer orbital 
periods will have
a higher average mass accretion rate (Verbunt \& van den Heuvel 1995). The correlation obtained by van den Heuvel \& 
Bitzaraki (1995) can therefore be explained by this dependence of the rate of accretion on orbital period. 

\section{conclusions}
\label{sconcl}

\ni  In this work, we have compared the outcome of the evolution of magnetic field located initially in the neutron 
star crust with field evolution trend observed in isolated as well as binary pulsars. We find that the model can 
explain almost all the features that have been observed to date. And our conclusions can be summarized as follows:
\bei
\i for this model to be consistent with the statistical analyses performed on the isolated pulsars a maximum value 
of 0.05 for the impurity strength can be allowed; 
\i HMXBs produce high-field long period pulsars provided the duration of the wind accretion phase is short
or the initial current distribution is located at higher densities;
\i relatively low-field ($B \sim 10^{10}$~Gauss) objects near death-line (low-luminosity pulsars) are also 
predicted from HMXBs;
\i LMXBs will produce both high-field long period pulsars as well as low-field short period pulsars inclusive
of millisecond pulsars in the later variety; and
\i a positive correlation between the rate of accretion and the final field strength is indicated and is 
supported by observational evidence.
\eei

\section*{Acknowledgments}
We thank Bhaskar Datta for supplying us with the data for a part of the equation of state. 
SK thanks R. Ramachandran, C. Indrani and Dipanjan Mitra for useful discussions.

\beb
\bi Alpar M.~A., Cheng A.~F., Ruderman M.~A., Shaham J., 1982, Nat, 300, 728
\bi Baym G., Pethick C., Sutherland P., 1971, ApJ, 170, 299
\bi Bailes M., 1996, {\em Pulsars : Problems and Progress}, ed. Johnston S.,
Walker M.~A., Bailes M., ASP Conference Series Vol.105
\bi Bhattacharya D., 1995a, JA\&A, 16, 217
\bi Bhattacharya D., 1995b, {\em X-Ray Binaries}, ed. Lewin W.~H.~G., van Paradijs J., 
van den Heuvel E.~P.~J., Cambridge University Press, pp. 233-251
\bi Bhattacharya D., 1996a, {\em Pulsars : Problems and Progress}, ed. 
Johnston S., Walker M.~A., Bailes M., ASP Conference Series Vol.105, pp. 547-556
\bi Bhattacharya D., 1996b, {\em Pulsar Timing, General relativity and 
the Internal Structure of Neutron Stars}, Royal Netherlands Academy of Arts and Sciences, in press.
\bi Bhattacharya D., Datta B., 1996, MNRAS, 282, 1059
\bi Bhattacharya D., Srinivasan G., 1986, Curr. Sci., 55, 327 
\bi Bhattacharya D., Srinivasan G., 1995, {\em X-Ray Binaries}, ed. 
Lewin W.~H.~G., van Paradijs J., van den Heuvel E.~P.~J., Cambridge University Press, pp. 495-522
\bi Bhattacharya D., van den Heuvel E.~P.~J., 1991, Phys. Rep., 203, 1
\bi Bhattacharya D., Wijers R.~A.~M.~J., Hartman J.~W., Verbunt F., 1992, A\&A, 254, 198
\bi Camilo O.~F., Nice D.~J., Shrauner J.~A., Taylor J.~H., 1996, ApJ, 469, 819
\bi Chen K., Ruderman M., 1993, ApJ, 408, 179
\bi Cordes J.~M., Chernoff D.~F., 1997, ApJ, 482, 971
\bi Cordes J.~M., Wolzscan A., Dewey R.~J., Blaskiewicz M., Stinebring D.~R., 1990, ApJ, 349, 245
\bi Geppert U., Urpin V.~A., 1994, MNRAS, 271, 490
\bi Hansen B.~M.~S., Phinney E.~S., 1998, MNRAS, 294, 569
\bi Hartman J.~W., Verbunt F., Bhattacharya D., Wijers R.~A.~M.~J., 1997, A\&A, 322, 477
\bi Hasinger G., van der Klis M., 1989, A\&A, 225, 79
\bi Jahan Miri M., Bhattacharya D., 1994, MNRAS, 269, 455 
\bi King A., Frank J., Kolb U., Ritter H., 1995, ApJ, 444, 37
\bi Konar S., Bhattacharya D., 1997, MNRAS, 284, 311
\bi Kulkarni S.~R., 1986, ApJ, 306, L85
\bi Kulkarni S.~R., Narayan R., 1988, ApJ, 335, 755
\bi Lorimer D.~R., 1994, {\em The Galactic Population of Millisecond and Normal Pulsars}, Ph. D. thesis, 
University of Manchester, U.K.
\bi Lorimer D.~R., 1995, MNRAS, 274, 300
\bi Negele J.~W., Vautherin D., 1973, Nucl.Phys., A207, 298
\bi Nicastro L., Johnston S., 1995, MNRAS, 273, 122
\bi Nice D.~J., Taylor J.~H., 1995, ApJ, 441,429
\bi Page D., 1998, {\em The Many Faces of Neutron Stars}, 
ed. Buccheri R., van Paradis J., Alpar M.~A., Kluwer Academic Publishers, Dordrecht, pp. 539-551
\bi Psaltis D., Lamb F.~K., 1997, preprint
\bi Ramachandran, R., Bhattacharya, D., 1997, MNRAS, 288, 565
\bi Romani R.~W., 1993, {\em Isolated Pulsars}, ed. van Riper K., Epstein R., Ho C., Cambridge University Press, pp. 75-82
\bi Ruderman M., 1995, JA\&A, 16, 207
\bi Sang Y., Chanmugam G., 1987, ApJ, 323, L61
\bi Urpin V.~A., Geppert U., 1995, MNRAS, 275, 1117
\bi Urpin V.~A., Geppert U., 1996, MNRAS, 278, 471 
\bi Urpin V.~A., Konenkov D., Geppert U., 1998, MNRAS, 299, 73
\bi Urpin V.~A., Muslimov A.~G., 1992, MNRAS, 256, 261
\bi van den Heuvel E.~P.~J., 1992, {\em X-ray Binaries and Recycled Pulsars},
ed. van den Heuvel E.~P.~J., Rappaport S., Kluwer academic Publishers, Dordrecht
\bi van den Heuvel E.~P.~J., 1995, JA\&A, 16, 255
\bi van den Heuvel E.~P.~J., van Paradijs J.~A., Taam R.~E., 1986, Nat, 322, 153
\bi van den Heuvel E.~P.~J., Bitzaraki O, 1995, A\&A, 297, L41
\bi Verbunt F., 1993, ARAA, 31, pp. 93-127
\bi Verbunt F., van den Heuvel E.~P.~J., 1995, {\em X-Ray Binaries}, ed. 
Lewin W.~H.~G., van Paradijs J., van den Heuvel E.~P.~J., Cambridge University Press, pp. 457-494
\bi Verbunt F., Wijers R.~A.~M.~J., Burm H., 1990, A\&A, 234, 195
\bi Wakatsuki S., Hikita A., Sato N., Itoh N., 1992, ApJ, 392, 628
\bi Webbink R.~F., Rappaport S., Savonije G.~J., 1983, ApJ, 270, 678
\bi Wijnands R., van der Klis M., 1998, Nat., 394, 344
\bi Wiringa R.~B., Fiks V., Fabrocini A., 1988, Phys.Rev., C38, 1010
\bi White N.~E., Zhang C.~M., 1997, ApJL, 490, 87
\bi Wolszcan A., 1994, Science, 264, 538
\eeb
\end{document}